\pdfoutput=1 
 \documentclass[%
 reprint,
 amsmath,amssymb,
 aps,
pre,
]{revtex4-1}

\usepackage{graphicx}
\usepackage{dcolumn}
\usepackage{bm}
\usepackage{subeqnar}
\usepackage{overpic}
\usepackage{color}
\usepackage{adjustbox}
\usepackage[caption=false]{subfig}
\usepackage{rotating}

\begin{document}

\title{Mode-Locked Rotating Detonation Waves: \\ Experiments and a Model Equation}

\author{James Koch}
\email{jvkoch@uw.edu}
 \affiliation{William E. Boeing Department of Aeronautics and Astronautics, University of Washington, Seattle 98195-2400}
  \author{Mitsuru Kurosaka}
 \affiliation{William E. Boeing Department of Aeronautics and Astronautics, University of Washington, Seattle 98195-2400}
 \author{Carl Knowlen}
 \affiliation{William E. Boeing Department of Aeronautics and Astronautics, University of Washington, Seattle 98195-2400}
\author{J. Nathan Kutz}
 \affiliation{Department of Applied Mathematics, University of Washington, Seattle, WA 98195-3925}

\begin{abstract}
Direct observation of a Rotating Detonation Engine combustion chamber has enabled the extraction of the kinematics of its detonation waves. These records exhibit a rich set of instabilities and bifurcations arising from the interaction of coherent wave fronts and global gain dynamics. We develop a model of the observed dynamics by recasting the Majda detonation analog as an autowave. The solution fronts become attractors of the engine; i.e., mode-locked rotating detonation waves. We find that denotative energy release competes with dissipation and gain recovery to produce the observed dynamics and a bifurcation structure common to driven-dissipative systems, such as mode-locked lasers.
\end{abstract}
\maketitle

\section{INTRODUCTION} \label{intro}
The {\em Rotating Detonation Engine} (RDE) is a thrust-producing device in which self-sustained combustion-driven shock waves, or detonations, travel azimuthally in an annular combustion chamber. Pressure rises through the detonation process, contrasting conventional deflagration-based engines. Successful implementation of so-called `pressure gain' combustion implies mechanical simplification of propulsion systems (for example, pumping requirements for propellant can be reduced \cite{Rankin2017}) and an increase of available work for a given propellant over conventional engines \cite{Nordeen2014}, ultimately resulting in fuel savings.  However, a diverse set of experimentally observed instabilities and bifurcations are known to be ubiquitous in  RDEs~\cite{Anand2015,Anand2017,Bennewitz2019}, potentially compromising performance and stable operation.  In this article, we develop a modeling framework that characterizes the  underlying global bifurcation structure of RDEs, showing that the nonlinear dynamics are governed by the interaction physics of global gain (fuel) depletion and recovery along with local dominant balance physics characterized by the Burgers' equation~ \cite{Majda1981}. Our predictions capture the cascade of bifurcations and flame-front solutions whose attracting nature we term {\em mode-locked rotating detonation waves} and which are observed experimentally within the RDE.  Further, the model shows that the underlying energy balance physics of the driven-dissipative RDE mimics those of mode-locked lasers~\cite{Haus2000mode,Kutz2006}, where global gain dynamics produce a similar cascading bifurcation diagram of mode-locked states~\cite{li2010geometrical}.

\begin{figure}[t]
        \centering
        \begin{overpic}[width=0.75\linewidth]{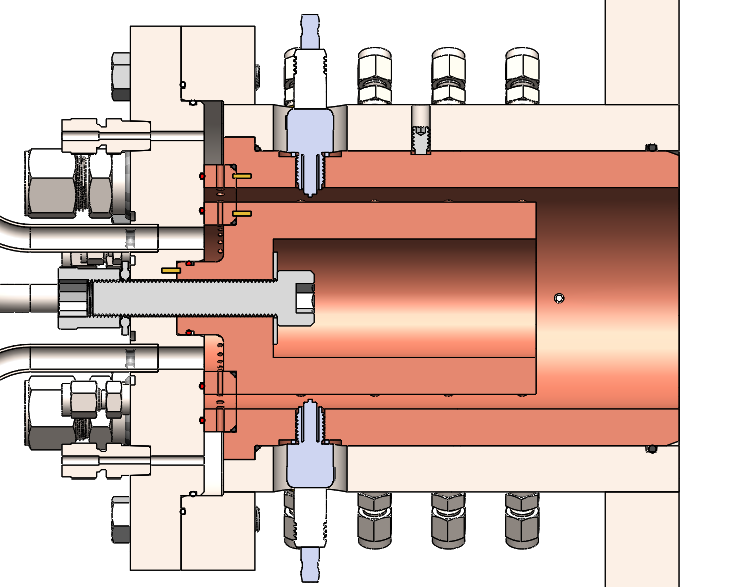}            
	    \end{overpic}  
		\caption{Section view of the Rotating Detonation Engine (RDE) used for this study. The engine geometry is such that gaseous methane and oxygen is directed into a narrow annular gap through a set of propellant injectors. A spark plug ignites the mixture, which rapidly transitions to a number of circumferentially traveling detonation waves.}
		\label{fig:cutaway}
\end{figure}

\begin{figure*}[t]
    \centering
    \begin{tabular}{cc}
    \adjustbox{valign=t}{\subfloat[\label{fig:frame}]{%
          
          \begin{overpic}[width=0.3\linewidth]{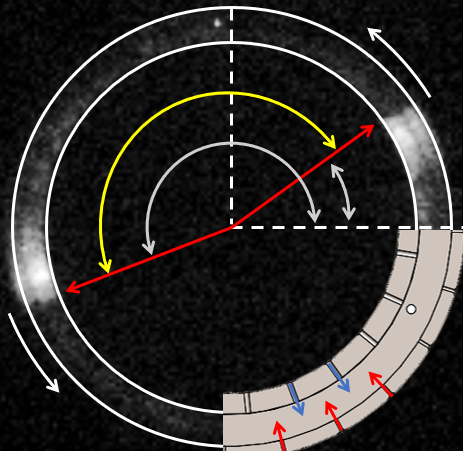}
		  \put(42,79){\color{yellow}{$\Psi=173^{\circ}$}}
		  \put(65,41){\color{white}{$\Theta=0^{\circ}$}}
  		  \put(79,50){\color{white}{$\Theta_1$}}
  		  \put(92,80){\color{white}{$\dot{\Theta}_1$}}
  		  \put(4,8){\color{white}{$\dot{\Theta}_2$}}
		  \put(35,47){\color{white}{$\Theta_2$}}
		  \put(55,22.5){\color{white}{O\textsubscript{2} Inj.}}
		  \put(78,1.5){\color{white}{CH\textsubscript{4} Inj.}}
		  \end{overpic}}}
		  
    & 
    \adjustbox{valign=t}{\begin{tabular}{@{}c@{}}
    \subfloat{

          \begin{overpic}[width=0.57\linewidth]{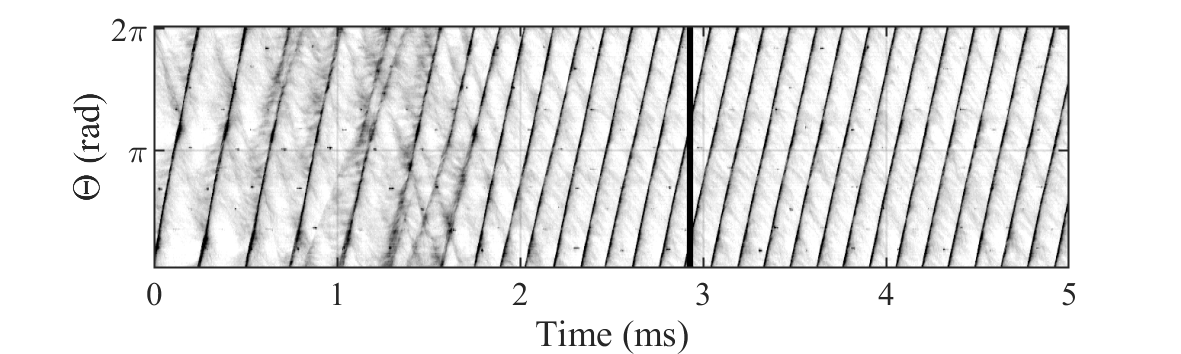}
		  \put(13,8){{(b)}}
		  \end{overpic}
		  \label{fig:xt}}
			\\[-.15in]
    \subfloat{
   
          \begin{overpic}[width=0.57\linewidth]{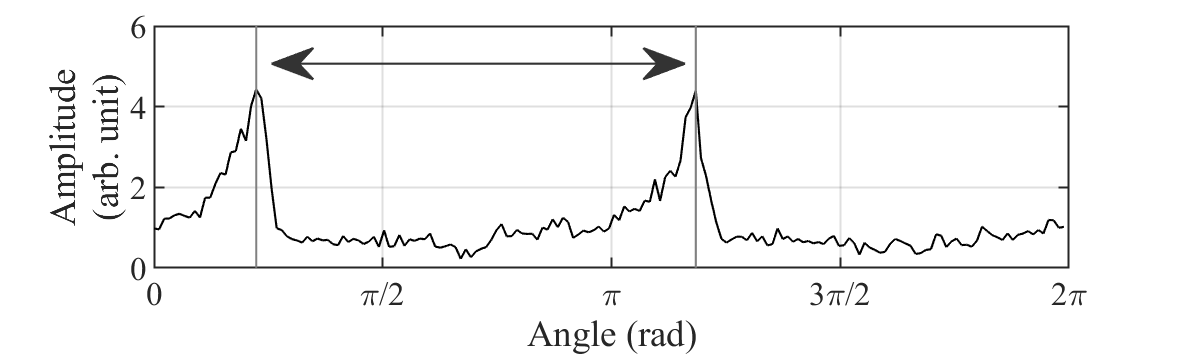}
		  \put(32,21){$\Psi= 173^{\circ}$}
		  \put(13,25){(c)}
		  \end{overpic}
		  \label{fig:domain}}
          
    \end{tabular}}
    \end{tabular}
    \caption{(a) A high-speed camera frame from an experiment shows the location of rotating detonation waves in the annulus of a RDE. Overlaid is a rendering of the propellant injection scheme. (b) Tracking the detonations through time yields a spatial-temporal view of their kinematics. Line slopes correspond to speed. The vertical cut in (b) is synchronized with the states of (a) and (c). (c) The phase difference $\Psi$ for the waves seen in (a) is not $\pi$, though eventually the phase difference approaches this stable value. }\label{fig:intro}
\end{figure*}
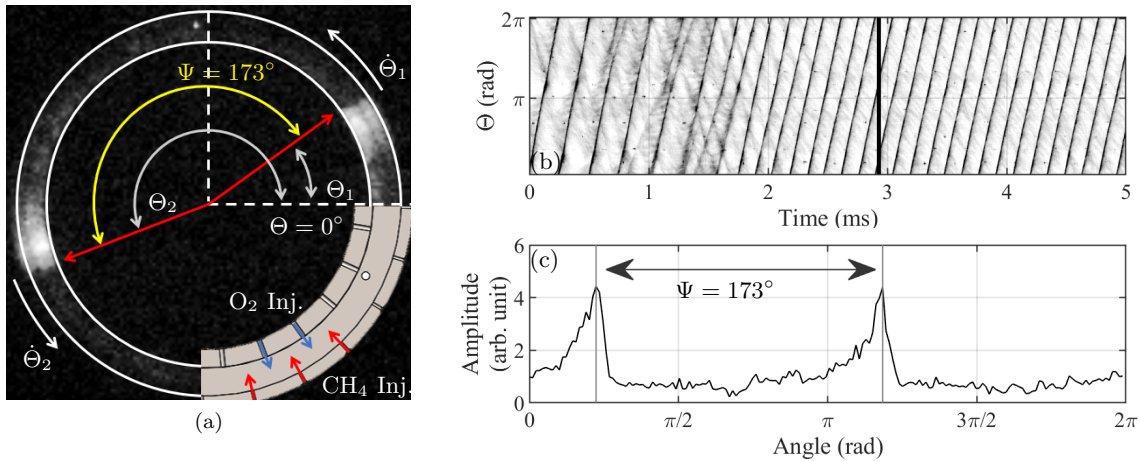

Conventional RDEs use concentric cylinders to direct the flow of propellant into a narrow annular gap (see Fig. \ref{fig:cutaway}). Inside this gap, an igniter deposits concentrated energy into the propellant mixture, creating an ignition kernel that promotes the exothermal chemical reaction. By virtue of the narrow annular gap, the gradients in density and pressure caused by the heat release self-steepen, eventually forming shocks strong enough to auto-ignite the propellant. These combustion-driven shock waves, now detonations, continue to process propellant so long as there is sufficiently fast refill and mixing of propellant within the period of the traveling detonation wave to offset inhibiting phenomena \cite{CULLEN1966,Bykovskii2006}. In this manner, the steady operation of the RDE is the point at which the rates of gain depletion (combustion), gain recovery (injection), and dissipation balance. For these to exist in an unbalanced state induces a degree of unsteadiness, typically manifested as a transition to a different number of waves or modulation of wave speed \cite{Anand2015,Anand2017,Bennewitz2019,Bohon2019}; i.e., the system bifurcates. 

In laboratory experiments, typical observables are wave count, speed, and direction as captured by pressure sensors or optical instruments. Common in experimental literature are a few themes: (i) the observed detonation wave speeds are significantly less than the Chapman-Jouget (CJ) velocity (the steady shock-induced combustion wave in which the combustion products are sonic relative to the wave front) for the propellant mixture \cite{Liu2015,Anand2019}, (ii) the number of waves is tied to the mass flow rate of the engine and the propellant injection scheme \cite{Bykovskii2006,Prakash2019}, and (iii) para-wave combustion, meaning deflagration not associated with a traveling wave, is ubiquitous \cite{Bohon2019,Bennewitz2019,Anand2019}. Additionally, we acknowledge the prevalence of counter-propagating waves in literature (see \cite{Bennewitz2019} and \cite{Anand2019}). However, for the present article we restrict our discussion to co-rotating waves only as a means to simplify the modeling and analysis.

Computational modeling of RDEs allows for detailed investigations of the wave structure and engine flowfield.  Not only do these models agree well with experiments, but they also produce many of the instabilities and observed bifurcations of RDEs, including mode-locked states~ \cite{Hishida2009,Shao2010,Schwer2011,Zhou2013}. However, these high-fidelity simulations are computationally expensive; i.e., to extract limit cycle behavior of the wave dynamics and bifurcation structures is not currently feasible.  Additionally, they fail to identify the leading order physics responsible for producing the bifurcations. Our modeling efforts draw on recent experimental observations of nonlinear dynamics of rotating detonation waves to formulate a reduced-order model description that captures the global bifurcations observed in practice. We have identified the dominant energy balance physics responsible for producing the universally observed physics of the mode-locked states and their interactions in many RDEs. Indeed, the primary bifurcation parameter controlling the cascade of bifurcations is easily identified as the propellant injection and mixing rate. The energy balance physics is canonical in that it is prevalent in a broader range of driven-dissipative physical systems, including  mode-locked lasers~\cite{Haus2000mode,Kutz2006,li2010geometrical,spaulding2002nonlinear}, Bose-Einstein condensates (BECs)~\cite{kutz2009mode}, and some biological systems \cite{2008a}.   Such rich bifurcation structures pervade spatio-temporal systems driven to instability~\cite{cross1993pattern}.

In Section \ref{section:experiments}, we describe the experimental apparatus and display recent observations of nonlinear dynamics within the engine. Building on these observations, a model system is proposed in Section \ref{section:model} with a goal of reproducing, qualitatively, the observed dynamics. Numerical experiments of the proposed model are presented in Section \ref{section:numerical} and follow with a discussion of the results in Section \ref{section:conclusion}.

\section{Experiments} \label{section:experiments}
For the present study, an RDE (Fig. \ref{fig:cutaway}) and test cell were designed and constructed to investigate rotating detonation wave dynamics. The engine used for this study is unique in that the engine internal components are modular. Engine cores can be swapped out to give different annular gaps and combustor lengths. The injectors can be similarly exchanged to investigate injector-combustion coupling and mixing strategies. The test cell is a backpressure controlled facility. Engine exhaust is routed to an appropriately sized vacuum chamber with a known backpressure. The test cell is optically accessible, which allows for recording the complete kinematic history of all detonation waves with high spatiotemporal resolution (Fig.~\ref{fig:frame}). Each experiment is a 0.5 second burn of a known proportion and feed rate of gaseous methane and oxygen. In a successful experiment, a spark ignites the mixture and produces an accelerating flame that transitions into a number of traveling detonation waves.  A complete description of the experimental apparatus and procedures are detailed in \cite{Boening2018}. 

\begin{figure}[t]
        \centering
        \begin{overpic}[width=1\linewidth]{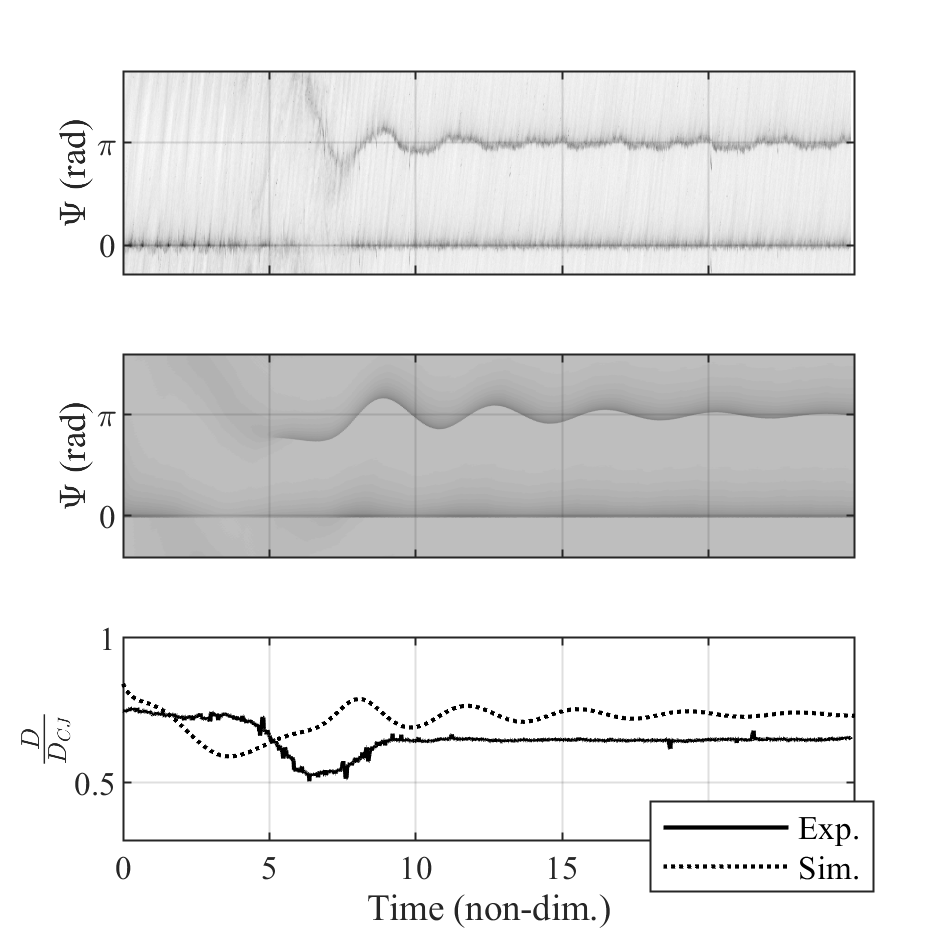}
        \put(14.5,88){{(a)}}	    
	    \put(14.5,58){{(b)}}    
	    \put(14.5,28){(c)}
	    \put(71,78){{Experiment}}
	    \put(78,48){{Model}} 	    
	    \end{overpic}
        \caption{Representative wave nucleation process in a startup transient in an experiment (a) and in a simulation of the proposed model (b) displayed as pseudocolor plots of amplitude (arb. units). As seen in the wave reference frame of (a) and (b), the oscillatory phase difference between the two waves immediately after nucleation decays through time as the two waves become mode-locked. (b) corresponds to $s = 3.5$. The wave speeds along $\Psi = 0$ in (a) and (b) are given in (c). }
		\label{fig:1to2}
\end{figure}

A fundamental assumption of this study is that the output luminosity in these experiments correlates to combustion progress, meaning brighter regions exhibit higher heat release than darker regions. Supposing this to be true, we examine example waveforms extracted from the high-speed camera footage. For each experiment, the azimuth-time history is extracted from high-speed video footage through a pixel-intensity integration algorithm \cite{Bennewitz2019a}. The wave kinematics can be recorded in this manner and displayed as a $\theta-t$ diagram, an example of which is shown in Fig.~\ref{fig:xt}. Furthermore, these records can be recast in the wave-attached frame, in which case the phase differences between waves is an explicit output (the tracked wave appears steady in this reference frame). Figure~\ref{fig:1to2}a is the data in Fig.~\ref{fig:xt} shifted to the wave reference frame. For these figures, we nondimensionalize time as $\tau = t  \left( D_{wave}/L \right)$, where $L$ is the length of the domain and $D_{wave}$ is the speed of the wave in its mode-locked state. 

In Fig.~\ref{fig:1to2}a, an observed transition from one wave to two waves during the startup transient is shown. In this mode transition, after a point of criticality, a second detonation wave forms and begins to travel around the annulus. However, the spacing between the two waves in the annulus is asymmetric, causing an imbalance in the amount of propellant consumed by each of the waves.  The wave with coordinate $\theta_1$ trailing the preceding wave $\theta_2$ exists with a phase difference of $\Psi = \theta_{2} - \theta_{1} < \pi$ (see Fig. \ref{fig:frame}). At that instant, assuming the propellant refresh rate is approximately constant, the trailing wave has less than half of available propellant in the chamber for its consumption. The local balance of gain (heat release), gain recovery, and dissipation is not satisfied. Since propellant heat release directly affects the speed of a detonation, the wave begins to decelerate. The preceding wave, however, has the remaining portion of available propellant and can accelerate through the excess of propellant. In this manner, these two waves behave dispersively, where they seek a stable state with maximum and symmetric phase differences. For the single wave portion in Fig.~\ref{fig:1to2}a, the quasi-steady wave has a velocity $20\%$ to $30\%$ below the Chapman-Jouget velocity for the propellant mixture. This metric is a direct observable of the energy necessary to sustain the detonation wave subject to dissipation and gain recovery in the combustion chamber. As the transition to two waves occurs and the dynamics settle to a steady state, the wave speed reduces to about $90\%$ of the single wave speed.

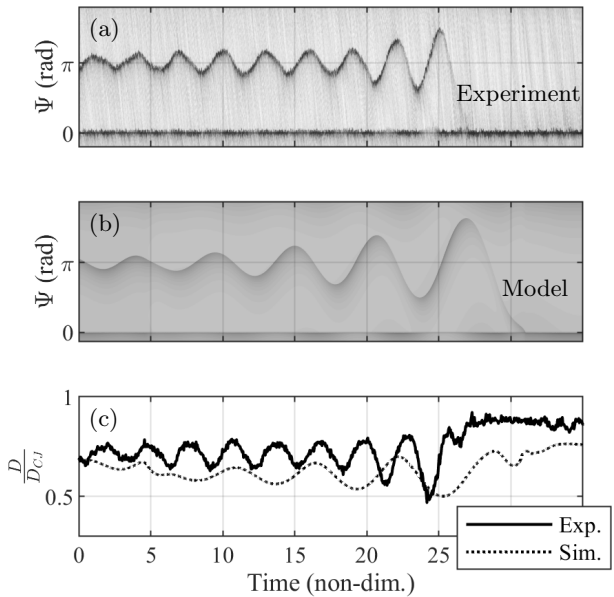
\begin{figure}[t]
        \centering
        \begin{overpic}[width=1\linewidth]{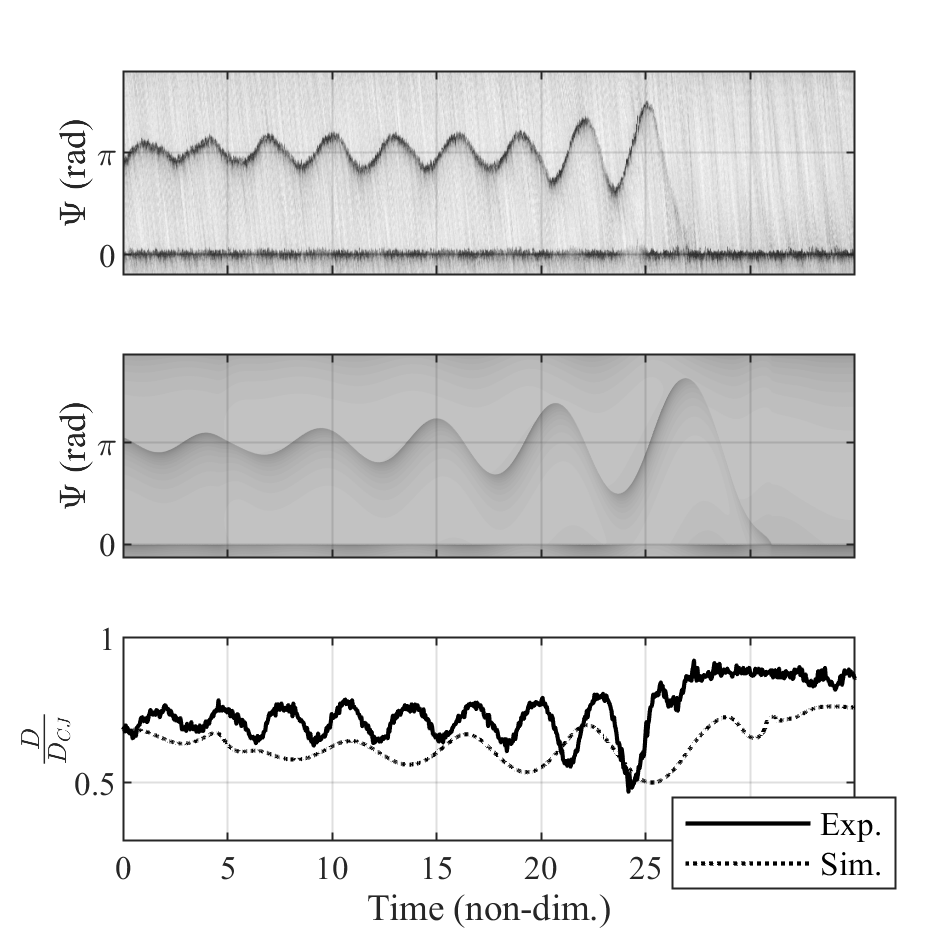}
        \put(14.5,88){{(a)}}	    
	    \put(14.5,58){{(b)}}    
	    \put(14.5,28){(c)}
	    \put(71,78){{Experiment}}
	    \put(78,48){{Model}}     
		\end{overpic}
        \caption{Representative destruction of a wave in an experiment (a) and in a simulation of the model (b) shown in the wave-attached reference frame as pseudocolor plots of amplitude (arb. units). Oscillations in $\Psi$ grow exponentially until one wave overruns the other. For a given injection function $\beta$ and loss $\epsilon$, the oscillation period and phase difference growth rate are parameterized by the change in $s$ and $u_p$. (b) corresponds to $s = 2$ with a $-20\%$ change in $s$ applied to the mode-locked state. The wave speeds along $\Psi = 0$ in (a) and (b) are given in (c).}
		\label{fig:transition}
\end{figure}

The opposite scenario occurs upon ramp-down of propellant feed at the end of each experiment. Figure \ref{fig:transition}a exemplifies a ramp-down transition of 2 to 1 wave over the span of about 10 ms. The two waves compete for the increasingly scarce propellant, as opposed to the case of excessive propellant exhibited in Fig. \ref{fig:1to2}a. Because of an initial perturbation in phase difference, the waves begin to exchange strength (speed and amplitude) in a regular fashion producing the exponential instability growth. As the phase difference oscillations grow, a catastrophic interaction between the waves occurs, resulting in the overrunning of the weaker wave by the stronger wave during one of the large-amplitude oscillations. After the bifurcation, the velocity of the remaining wave is about 10\% higher than that of the wave prior to the instability.

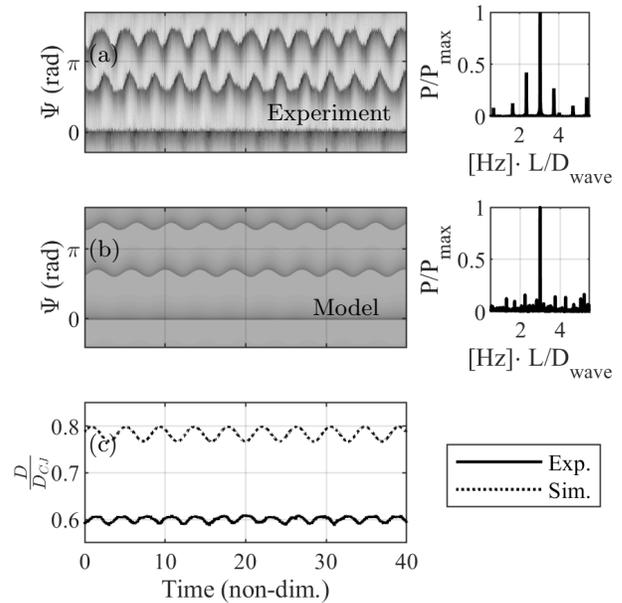
\begin{figure}[t]
        \centering
        \begin{overpic}[width=1\linewidth]{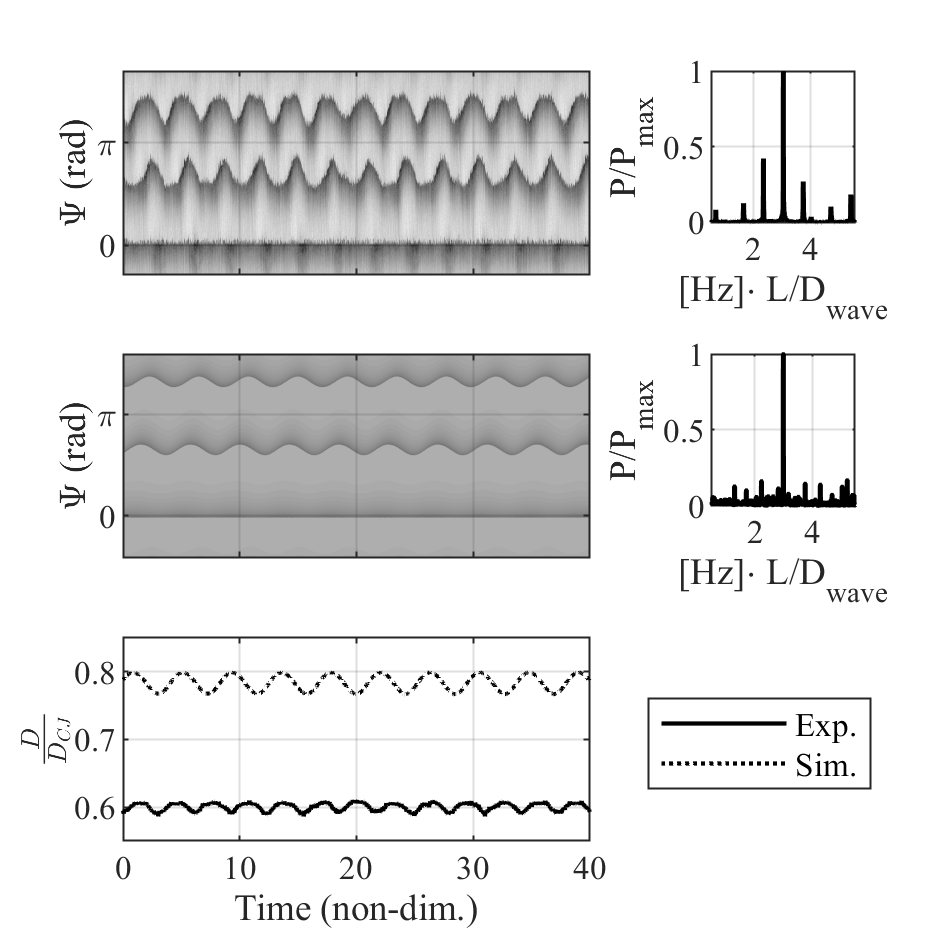}
        \put(13.5,85){{(a)}}	    
	    \put(13.5,55){{(b)}}    
	    \put(13.5,25){(c)}
	    \put(41,76){{Experiment}}
	    \put(48,46){{Model}}            
	    \end{overpic}  
		\caption{Space-time history of mode-locked modulation of wave speeds an experiment (a) and in a simulation (b) in the wave reference frame. Wave speeds along $\Psi=0$ in (a) and (b) are shown in (c). The accompanying spectra show clear sidebands symmetric about the carrier frequency.}
		\label{fig:3waveMod}
\end{figure}

Wave instabilities that do not lead to a change in the number of waves are common in the tested set of hardware. Fig. \ref{fig:3waveMod} exhibits a periodic wave velocity and amplitude observed in an experiment with three co-rotating waves. This is a clear modulational instability as spectral sidebands accompany the carrier frequency corresponding the the mean traveling wave velocity in the combustion chamber. This mode of operation is stable in the sense that it does not lead to a bifurcation of number of waves unless the flow condition is perturbed significantly.

\begin{figure}[t]
        \centering
        \begin{overpic}[width=1\linewidth]{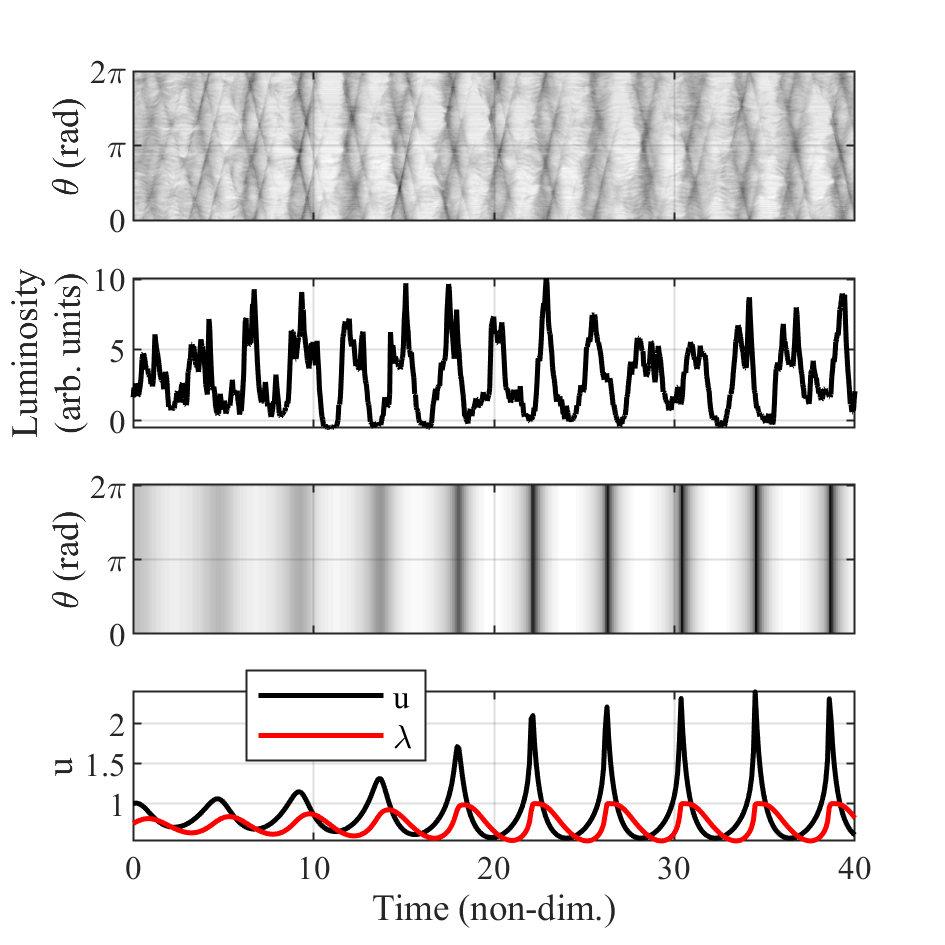}
        \put(14.5,89){(a)}	    
	    \put(14.5,67){(b)}    
	    \put(14.5,45){(c)} 
		\put(14.5,23){(d)} 	    
	    \end{overpic}  
		\caption{Space-time history of plane wave pulsation mode of operation in an experiment (a,b) and in a simulation (c,d). Simulation parameters are those listed in Table \ref{tab:simulations} with $q_0 = 6$ and $\epsilon = 1.0$. The deactivation and reactivation of the injectors gives rise to a resonance between the combustion and propellant injection.}
		\label{fig:planeWave}
\end{figure}
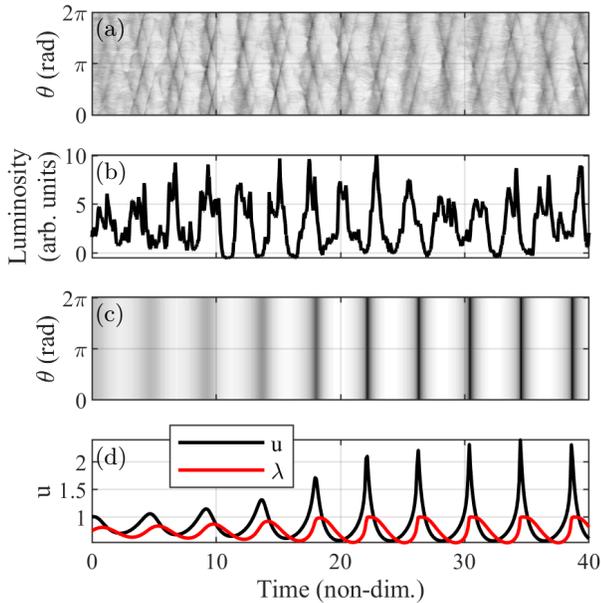

Pulsating modes of operation have also been observed in some experiments with very large injector areas (relative to the area of the annular combustion chamber). This mode of operation is characterized by a binary `on/off' behavior of the injectors and subsequently mixing and combustion. The oscillatory plane waves from an example pulsating mode is given in Fig. \ref{fig:planeWave}.

\section{A Qualitative Model} \label{section:model}

We propose a model that captures the dominant physics involved in the processes of wave formation, mode-locking, and mode bifurcations for further study of these phenomena. The detonation analogs of Majda \cite{Majda1981}, Fickett \cite{Fickett1985}, Rosales \cite{Rosales1983}, and Faria and Kasimov \cite{Faria2015} have enabled the rigorous mathematical description of detonation stability \cite{Lyng2004} and detonation dynamics in one (limit cycles and chaos) and two dimensions (cells and pattern formation). These analyses typically occur in the Lagrangian, shock-attached framework under assumptions of complete combustion. We use Majda's analog as a starting point as it sufficiently captures the dominant shock-chemistry interplay found in detonation waves.  Specifically, we recast Majda's analog in terms of autowave-producing variables \cite{Vasilextquotesingleev1979,Kerner1989}. Our model captures the dominant physics of gain depletion, gain recovery, and dissipation whose structure is given by:
\begin{subeqnarray}
&& \hspace*{-.4in} u_t + u u_x = \left( 1 - \lambda \right)\omega\left(u\right) q_0 + \nu u_{xx} + \epsilon \xi\left(u,u_0 \right)\\
&& \lambda_t = \left( 1 - \lambda \right)\omega\left(u\right) - \beta \left(u,u_p,s\right)\lambda
\end{subeqnarray}
where $u(x,t)$ is a quantity holding weak relationships to density and velocity (see \cite{Majda1981}) and $ \lambda$ is a combustion progress variable ($\lambda = 0$ is unburnt and $\lambda = 1$ is complete combustion). Gain is modeled with a heat release function, $\omega\left(u\right)$ with heat release $q_0$ as a proportionality constant. Dissipation and losses are modeled with a diffusion term $\nu u_{xx}$ and generic loss function $\epsilon \xi \left(u,u_0\right)$, where $\nu$ is viscosity, $u_0$ is the ambient state of the combustor, and $\epsilon$ is a loss magnitude constant. Lastly, the gain recovery is dictated by the injection model $\beta\left(u,u_p,s\right)$, where $u_p$ and $s$ are injection parameters. The domain is restricted to a 1-D periodic line in the Eulerian reference frame. 

The exact functional forms of the gain depletion, gain recovery, and loss terms are not critical to produce mode-locked rotating detonation waves. However, the inclusion of each of these terms in the model system is critical - to omit any one will destroy the balance required to provide the necessary properties and dynamics relevant to RDEs. In the opinion of the authors, presented herein are the simplest viable functional forms to provide the dynamics observed in real engines. These terms undoubtedly require modifications and/or parameter changes to mimic a specific set of hardware, but the underlying physical principles modeled by these terms are hypothesized to persist among all RDEs.

\subsubsection*{Gain Depletion}
The heat release function $\omega(u)$ is dictated by a simplified version of Arrhenius kinetics with a explicitly defined `ignition temperature' $u_c$ and activation energy $\alpha$: 

\begin{equation} \label{eqn:kinetics}
\omega\left(u\right) =  \exp \left( {\frac{u - u_c}{\alpha}} \right)
\end{equation}

For a steadily traveling detonation wave, the expectation is that this gain term dominates the dynamics, providing a rapid release of energy into the domain saturable only by exhaustion of fuel or another nonlinear effect (such as a nonlinear loss term). 

\subsubsection*{Losses and Dissipation}
The loss of energy in the domain is taken to be a generic restoring force to a natural state; i.e., the state of the of fresh propellant entering the combustion chamber ($u_0$). Physical mechanisms for loss include rejecting exhaust gases to an ambient condition and heat losses to the walls of the combustor. For model simplicity, we lump these effects into an assumed functional form:
\begin{equation}
\xi\left(u,u_0\right) = \left(u_0 - u\right) u^n
\end{equation}
This loss function is generic in that the relative significance of losses compared to gain can be modified by the proportionality constant $\epsilon$. Additionally, the linearity of the loss term can be prescribed with the index $n$. In this paper, we explore both linear ($n = 0$) and quadratic ($n = 1$) losses. For simplicity, we take $u_0 = 0$ such that the loss terms become $-\epsilon u $ or $-\epsilon u^2$ in the linear and quadratic loss cases, respectively. A diffusion term, $\nu u_{xx}$, is included in the model to retain the necessary generality for ongoing and future analysis.

\subsubsection*{Gain Recovery}
\begin{figure}[t]
        \centering
        \begin{overpic}[width=1\linewidth]{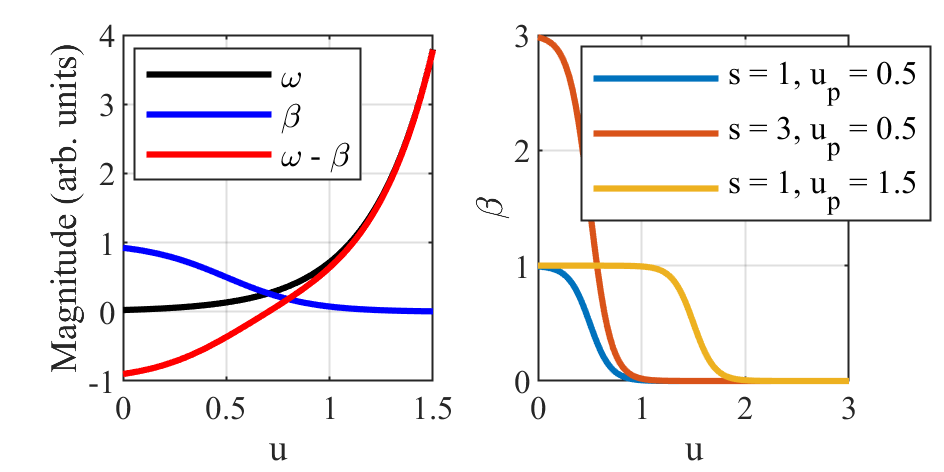}
	    \end{overpic}  
		\caption{The influence of the state of the domain on the balance between gain depletion and recovery (left) and recovery exclusively (right) following the functional form of Eqn. \ref{eqn:injection}.}
		\label{fig:injectors}
\end{figure}

The gain recovery term $\beta\left(u,u_p,s\right)$ works against gain depletion to `refill' the domain towards a $\lambda = 0$ state. In gaseous injection, injectors are typically `choked' orifices, meaning that perturbations in the combustor cannot influence the injection process as no characteristics can travel upstream past the choke point. However, in the presence of large-amplitude pressure oscillations (such as those present in detonation engines), the peak pressures may be comparable to those of the propellant plenums. This implies a loss of the sonic condition of the injectors. Should this occur, the state of the combustor becomes coupled to the injection scheme and can lead to unsteady behavior. In RDEs, the pressures generated by the detonation waves can be an order of magnitude larger than those of the propellant feed plenums. The injectors are periodically blocked (cutoff of injection) and backflow may be induced into the plenum chambers, further disrupting the injection process. To include these phenomena into the model, we use an activation function-based injector term that responds to the periodic forcing by the rotating detonation waves. The proposed activation function is given by:

\begin{equation} \label{eqn:injection}
\beta\left(u,s\right) = \frac{s}{1 + e^{k\left( u - u_p \right)}}
\end{equation}

where $s$ is a parameter analogous to injection area, $u_p$ is the injector `plenum pressure', and $k$ is a parameter adjusting the `steepness' of the activation function. Increasing injection area ($s$), plenum pressure ($u_p$), or both increases the mass flux into the engine. However, the dynamic response to these increases differs significantly. In the case of a high plenum pressure (a `stiff' injector), the influence of the detonation pressure becomes insignificant and the injector can deliver a consistent supply of propellant. In the case of a large injection area (holding the plenum pressure constant), the injectors are susceptible to large fluctuations of mass flux in response to this periodic forcing. Example activation function-based injector models are shown in Fig. \ref{fig:injectors}. Mixing is assumed to be exponential with time - in the absence of combustion, $\lambda$ asymptotically approaches 0. 

\begin{table} 
	\caption{Simulation Parameters}
	\label{tab:simulations}
	\centering
	\begin{tabular}{cccccccccccc}
	\hline
	$L$& $q_0$ &$\nu$ &$\alpha$ &$u_c$ &$u_0$ & $u_p$ & $k$& $\epsilon$ &$n$& $D_{CJ}$\\
	
	$2\pi$ & 1.0 & 0 & 0.3 & 1.1 & 0 & 0.5 & 5 & 0.11 & 1 &2\\
	\hline
	\end{tabular}
\end{table}

\begin{figure*}[t]
        \centering
        \begin{overpic}[width=1\linewidth]{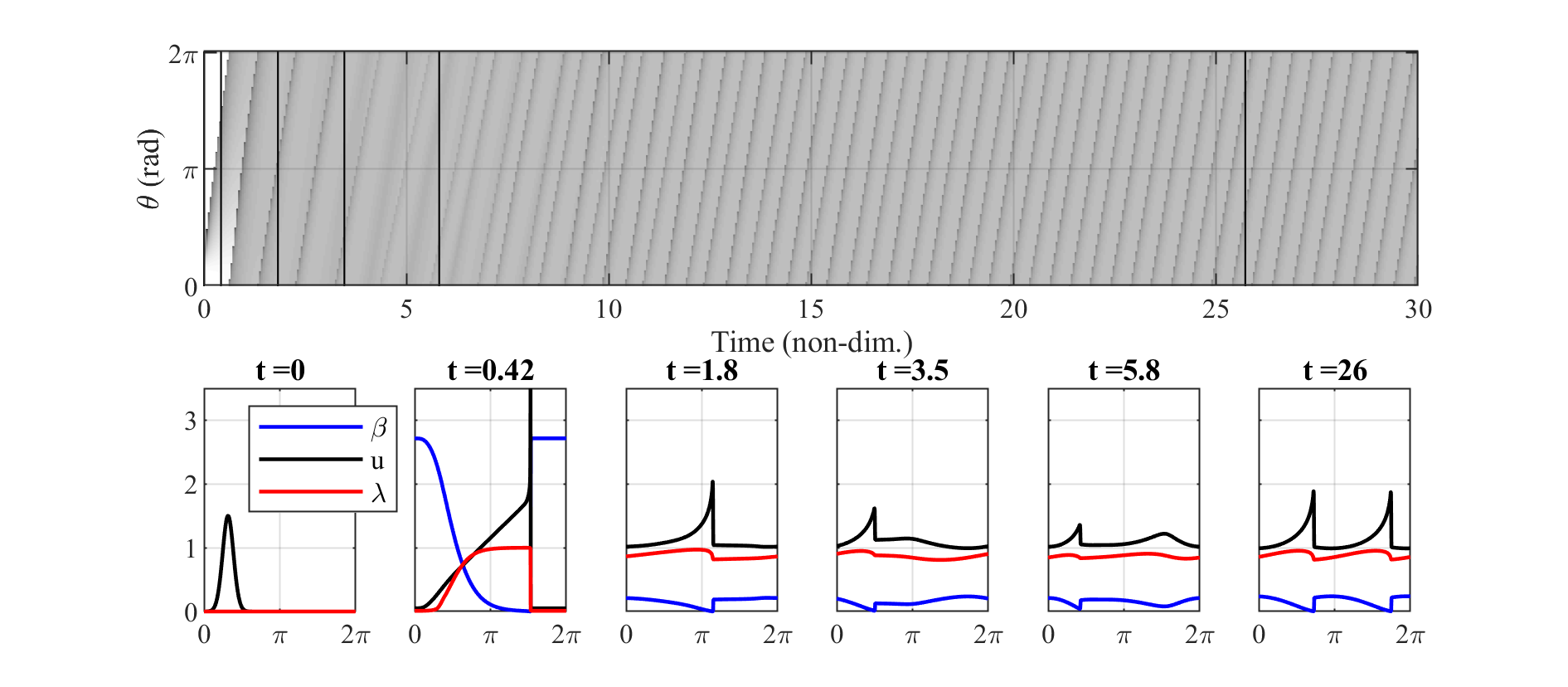}	   
	    \end{overpic}  
		\caption{Nucleation and mode locking of detonations from a single pulse initial condition ($s = 3.5$). Vertical lines in the $\theta-t$ diagram correspond to simulation snapshots shown. The initial \textit{sech}-pulse rapidly transitions to a CJ detonation. In regions where $u$ is low, the injectors behave steadily. However, as the wave reaches its tail, the $u$ is everywhere elevated and the injection is severely curtailed. A second wave forms from the self-steepening of para-wave deflagration. After wave nucleation, the two waves behave dispersively and their phase differences approach $\pi$.}
		\label{fig:pulses}
\end{figure*}

\section{Numerical Experiments} \label{section:numerical}
Numerical simulations are performed with the PyClaw open source finite volume software \cite{Ketcheson2012} on a converged grid. The parameters used for the numerical simulations in this article are given in Table \ref{tab:simulations}. Exceptions are noted as appropriate.

\subsubsection*{Planar Fronts}
We first examine the existence of planar solutions to the model system, including limit cycle behavior. The initial value problem was solved with initial condition $u(x,0) = \lambda(x,0) = 0.75$. A plane wave oscillates about the point in phase space where gain depletion and gain recovery match ($\beta\lambda = (1-\lambda)\omega$) subject to the balance of energy input and dissipation ($\epsilon\xi = (1-\lambda) \omega(u)q$). Low-energy oscillations decay to a planar deflagration front without oscillations. Pulsating fronts, such as those seen in recent experiments, are characterized by periodic `activation' and `deactivation' of the injectors - first resonating with the heat release, and subsequently saturated by the loss mechanisms. An example of a plane wave pulsating front can be seen in Figure \ref{fig:planeWave}d for a single location in the annulus through time. The corresponding space-time history for the pulsating mode of operation is given in Figure \ref{fig:planeWave}c. Pulsating plane wave solutions of the full model are stable for planar initial conditions, but are unstable to perturbations as they grow into traveling detonation waves.  

\subsubsection*{Traveling Waves}

For traveling wave simulations, the initial value problem with initial condition $u(x,0) = ({3}/{2})\mbox{sech}^{2} \left(x - x_o\right)$ and $\lambda(x,0) = 0$ was solved under varying refill ($s$, holding $u_p$ constant) conditions and with linear and nonlinear loss terms.

As in \citep{Majda1981}, we find the analogous CJ velocity of the reduced system (the inviscid, steady wave in which all energy has been released to the wave in a infinitesimally thin reaction zone). This steady wave speed is defined as the minimum speed that fulfills the Rankine-Hugoniot conditions for the prescribed heat release. In the limit as viscosity becomes zero and in the absence of losses, this minimum speed (CJ velocity) is $u_{CJ} = \left( u_1 + q_0\right) + \sqrt{q_0\left( q_0 + 2u_1\right)}$, where $u_1$ is the upstream state of a steady, shock-attached framework of the Majda Model. In the case of $u_1 = 0$, the speed of the CJ wave becomes $u_{CJ} = 2q$. This speed is the metric upon which the traveling waves in the proposed model are benchmarked. 

The evolution of a typical simulation is given in Fig. \ref{fig:pulses}. Because the initial $sech$-pulse is well above $u_c$, the medium locally and rapidly releases heat. The wave steepens and forms a detonation. This initial pulse travels at the CJ speed until it reaches its tail, at which point the wave begins to rapidly dissipate: the limited amount of gain recovery cannot sustain the CJ wave. Additionally, the rapid heat release (compared to the time scale of the dissipation of energy) of the initial CJ wave acts to raise the average $u$ in the domain substantially above the ambient value $u_0$ and ignition value $u_c$. In this manner, the \textit{effective} activation energy of the active medium is lowered and para-wave deflagration, or slow-scale heat release not associated with the traveling waves, is promoted in the entirety of the domain. Because the transit time of the initial traveling wave has been increased through dissipation, the formation of multiple, lower-amplitude pulses occurs by deflagration-to-detonation transition (DDT). 

To induce a mode transition from an already mode-locked state, a step change in $s$ is applied to the steady state, inducing a bifurcation. An example of such a transition is shown in Fig. \ref{fig:transition}b, where two initially mode-locked rotating detonation waves become unstable and destructively bifurcate. Low-amplitude phase difference oscillations grow exponentially, much like the experimental observations in Fig. \ref{fig:transition}a. During the period of oscillations, the two waves exchange strength (amplitude) and speed. For a given injection function $\beta$ and loss $\epsilon$, the instability growth rate and oscillation period is parameterized by the severity of the applied step in $s$ and $u_p$.

Upon nucleation of a new wave or destruction of an existing wave, the collection of waves in the chamber act dispersively, eventually forming a mode-locked state. The spatial imbalance of gain and dissipation in the domain allows for the characteristic modulation of detonation wave speed and amplitude. In transients of gain recovery, such as when the mass flow rate of an experiment is not constant, seen is a local imbalance of the gain and dissipation that either nucleates a new wave or amplifies asymmetric perturbations between waves, eventually causing a catastrophic destructive interaction.

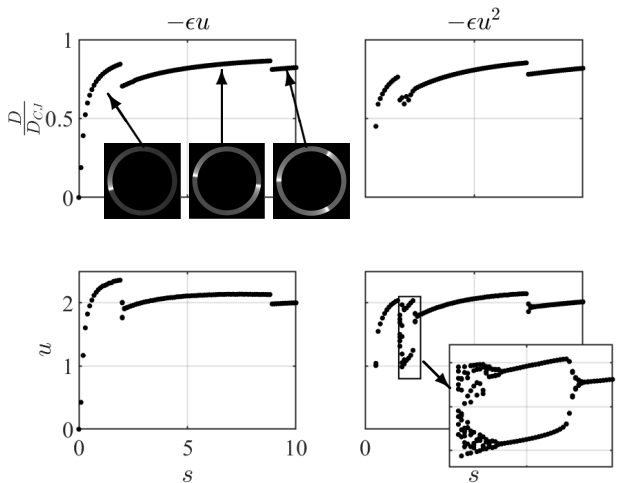
\begin{figure}[t]
        \centering
		\begin{overpic}[width=1\linewidth]{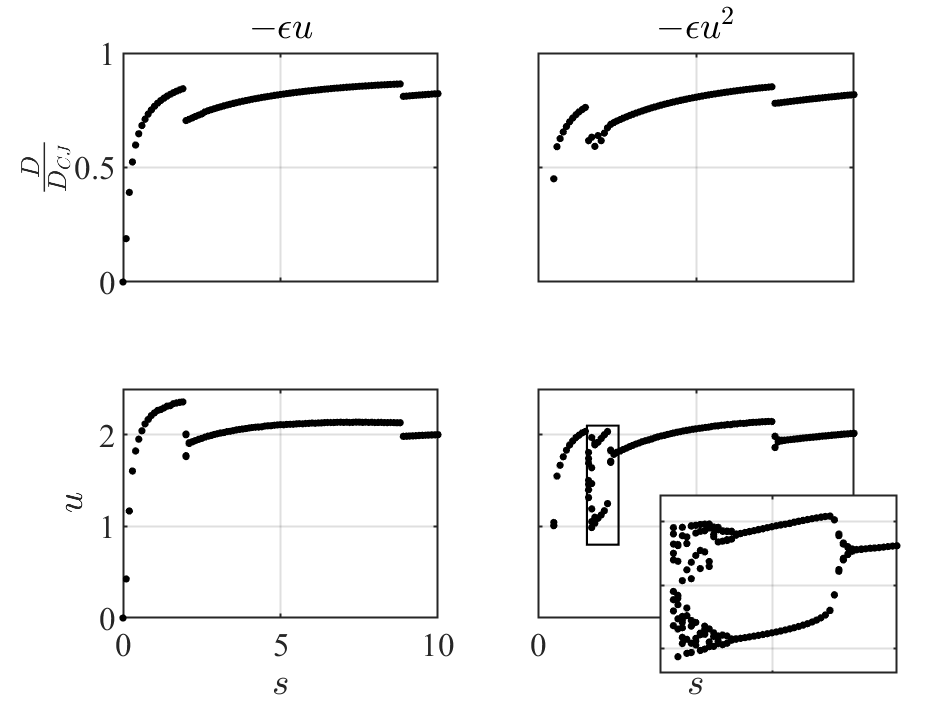}
     	\put(17,42){\includegraphics[scale=0.11]{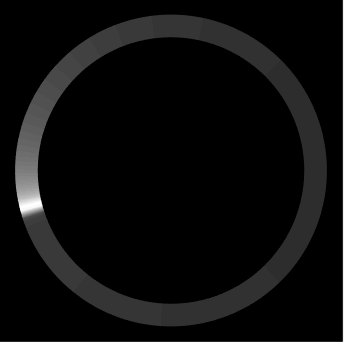}}
     	\thicklines
		\put(23,53){\vector(-2,3){5.5}}
     	
     	\put(30,42){\includegraphics[scale=0.11]{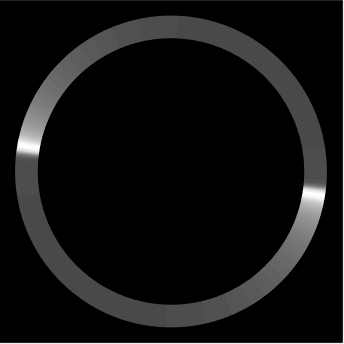}}
     	\put(35,53){\vector(0,1){12}}
     	 
     	\put(43,42){\includegraphics[scale=0.11]{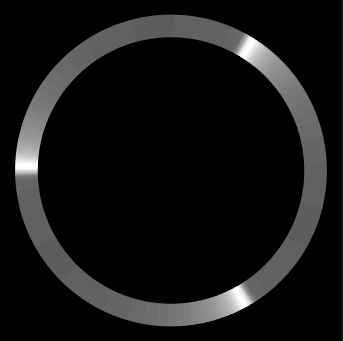}}
     	\put(48,53){\vector(-1,4){3}}

		\put(66,20){\vector(1,-1){4.1}}		
		\end{overpic}
		  \caption{Number of waves and wave speed through a sweep of the bifurcation parameter $s$ for linear and nonlinear loss terms. The final states have been approached from `above'; i.e., via a relaxation of three waves to two or one waves. In the transition from one to two waves for the nonlinear loss case, a series of period-halving bifurcations increase order in the system to eventually form two mode-locked waves with zero oscillation in phase difference. The qualitatively similar experimental bifurcation structure is published in Bykovskii et al. \cite{Bykovskii2006}.  The energy balance dynamics in laser cavities  produce a similar cascaded bifurcation structure, including chaotic inter-pulse regimes~\cite{li2010geometrical,bale2009transition}.}
		\label{fig:modes}
\end{figure}

Bifurcation diagrams showing the dependence of number of waves, wave speed, and wave amplitude on $s$ and the loss term is shown in Fig. \ref{fig:modes} for the parameters of Table \ref{tab:simulations}.  As $s$ is increased from zero, steady planar deflagration fronts form for small values. Once the value of $s$ can support a traveling wave, the waves follow the staircase behavior in Fig. \ref{fig:modes}, where the wave speed increases until another bifurcation occurs. These waves nucleate from the para-wave deflagration through a DDT process (an example of which is shown in Fig. \ref{fig:1to2}b). At each bifurcation to an increased number of waves, there is a drop in wave speed, though this drop in speed becomes less severe as the number of waves increases. This phenomena is consistent with the presented experiments as well as the observations of many in the literature \citep{Bykovskii2006}. As $s$ is further increased, the number of waves increases until the wave fronts are low in amplitude and merge into a planar deflagration front. 

For sweeps of the bifurcation parameter $s$ with an imposed quadratic loss, a series of period-halving bifurcations increase order in the system during the transition from one to two waves (Fig. \ref{fig:modes}d). In the regime of chaotic propagation, there is aperiodic nucleation, destruction, and modulation of the waves. As the gain is increased, the waves transition to periodic modulation of wave speed and phase difference. This characterisitic modulation is also seen in the transition from two to three waves (as in Fig. \ref{fig:3waveMod}). These intermediate modes are stable (persist for long durations). A significant degree of hysteresis is also noted in the regions near mode changes. Approaching criticality for any bifurcation-inducing parameter from above or below gives different behavior near the bifurcation. For example, a portion of the chaotic region in Fig. \ref{fig:modes} exhibits single-wave and dual wave chaotic multistability depending on single or two or more wave initial conditions. 

\section{Discussion} \label{section:discussion}
The analog system presented in this paper qualitatively reproduces the nonlinear dynamics of collections of waves observed in experiments. The proposed system is an adaptation of the Majda detonation analog to a periodic domain with gain depletion, gain recovery, and generic restoring forces included in the system. These terms sufficiently mimic real-engine processes such as heat release, propellant injection, and expulsion of exhaust to an ambient condition. While we have not explicitly captured all physical processes involved in real engines, nor have we perfectly identified the functional forms for the included terms, we do claim to have identified the dominant balance physics involved in the nonlinear dynamical behavior seen in real engines. These phenomena include wave nucleation (Fig. \ref{fig:1to2}), mode locking of multiple waves (Fig. \ref{fig:pulses}), wave destruction (Fig. \ref{fig:transition}), wave modulation (Fig. \ref{fig:3waveMod}), and pulsating plane waves (Fig. \ref{fig:planeWave}).

\subsection{Communication Pathways}
In steady operation of an RDE and in the mode-locked state of the proposed model system, a number of traveling detonation waves co-exist in the periodic domain with maximum possible phase differences between the waves. Supposing these traveling waves to be detonations, this would imply a lack of communication between the waves: detonations travel supersonically and, if steady, in a condition where the combustion products are sonic relative to the wave front. For the waves to behave dispersively, as in Figs. \ref{fig:1to2} and \ref{fig:transition} near the bifurcation points, implies a significant communication pathway or coupling mechanism. This mechanism is through the injection scheme and subsequently properties of the medium through which the detonations propagate. The injection scheme is responsible for providing a consistent combustible medium through which the detonations can propagate. However, known is that detonations induce blockages or backflow into propellant plenums. This phenomena is captured in our proposed functional form of $\beta$ in Eqn. \ref{eqn:injection}, providing a necessary feedback mechanism between the detonations and the injection scheme. In this manner, the presence of all detonation waves is impressed upon the dynamic response of the injectors and long-range communication is established, allowing for dispersive behavior. We therefore conclude that the coupling of the injectors and the detonation waves is what drives the observed dynamics in both experiments and in the proposed model, subject to the constraint of the generic losses inside of the chamber. In the presence of this non-locality, domain periodicity, and nonlinear gain and loss terms, chaotic solutions have been found to exist, as shown in Fig. \ref{fig:modes}.

\subsection{Bifurcations}

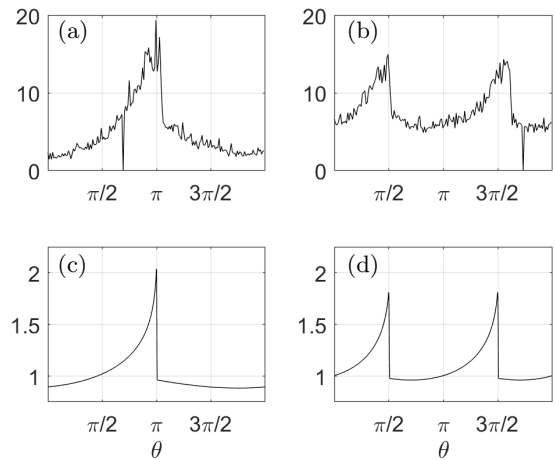
\begin{figure}[t]
        \centering
		\begin{overpic}[width=1\linewidth]{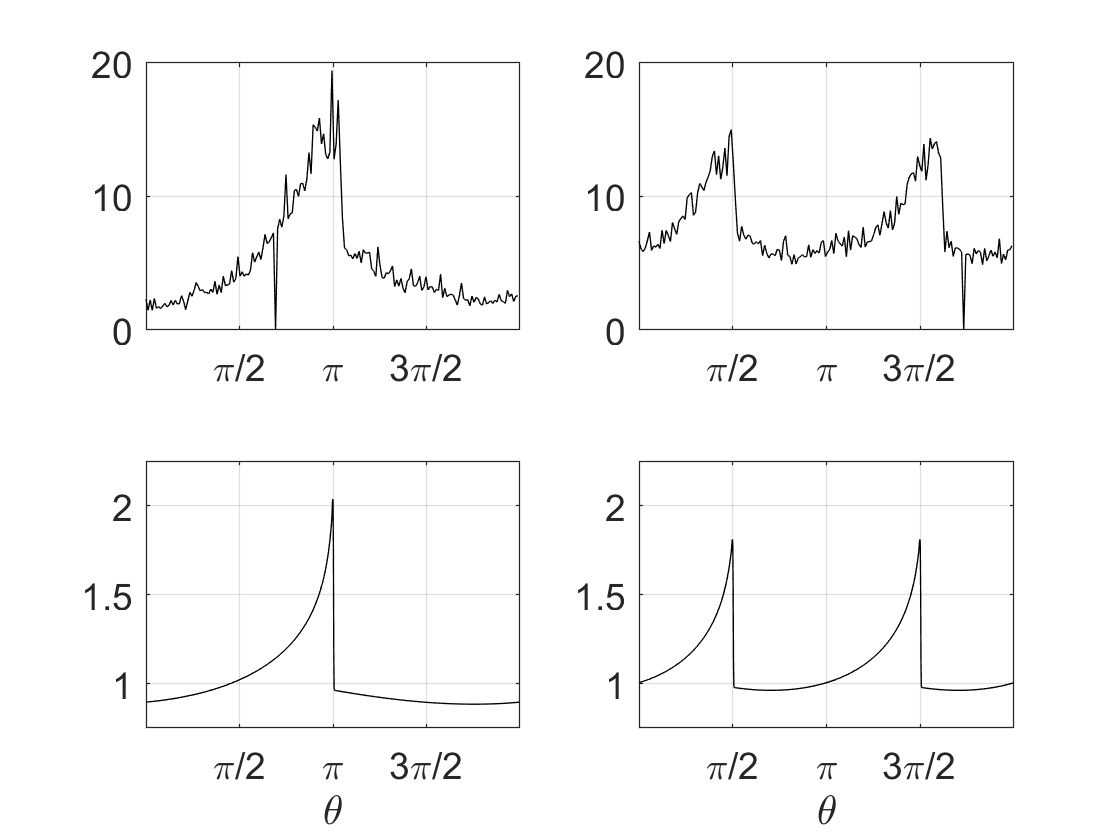}
		\put(14.5,66){(a)}	    
	    \put(59,66){(b)}    
	    \put(14.5,30){(c)} 
		\put(59,30){(d)} 	
		\end{overpic}
		  \caption{Integrated pixel intensity displayed through $\theta$ for an experiment through which a mode transition from one (a) to two (b) waves occurs. A similar induced bifurcation in a simulation is displayed in (c) and (d). Of note is (i) the decrease of wave amplitude between one and two waves, (ii) the increase in background luminosity (or, in the simulation, base state of $u$), and (iii) the local increase of the magnitude of the state preceding the waves. The wave speeds decrease about 10\% through the bifurcation, though this decrease is attributable to both a reduction in wave amplitude and an increase in para-wave deflagration. Once the para-wave deflagration can self-steepen to form a shock (see Fig. \ref{fig:pulses}), a bifurcation of number of waves occurs.}
		\label{fig:waveform}
\end{figure}

The presence of para-wave deflagration and weak restoring forces are the key physical mechanisms identified in the model system for inducing bifurcations. Within the model system, the time scale for detonative energy release is significantly shorter than those of deflagration and the generic losses. Therefore, at the onset of detonation, there is local accumulation of energy that will take a significant amount of time to dissipate to return the domain to a natural state (longer than the time-of-flight of a traveling wave). However, gain depletion is governed by simplified Arrhenius kinetics (Eqn. \ref{eqn:kinetics}), where the state of the domain $u$ is now elevated because of the slow-scale energy dissipation. In effect, the weak restoring force acts to promote kinetics in the chamber. Analogous physical mechanisms in real engines include preheating of the propellant and insufficient expulsion of burnt propellant from the combustion chamber, leading to an increase of temperature in the domain and subsequently faster kinetics. This results in an increased susceptibility to para-wave deflagration. 

\begin{figure}[t]
        \centering
		\begin{overpic}[width=1\linewidth]{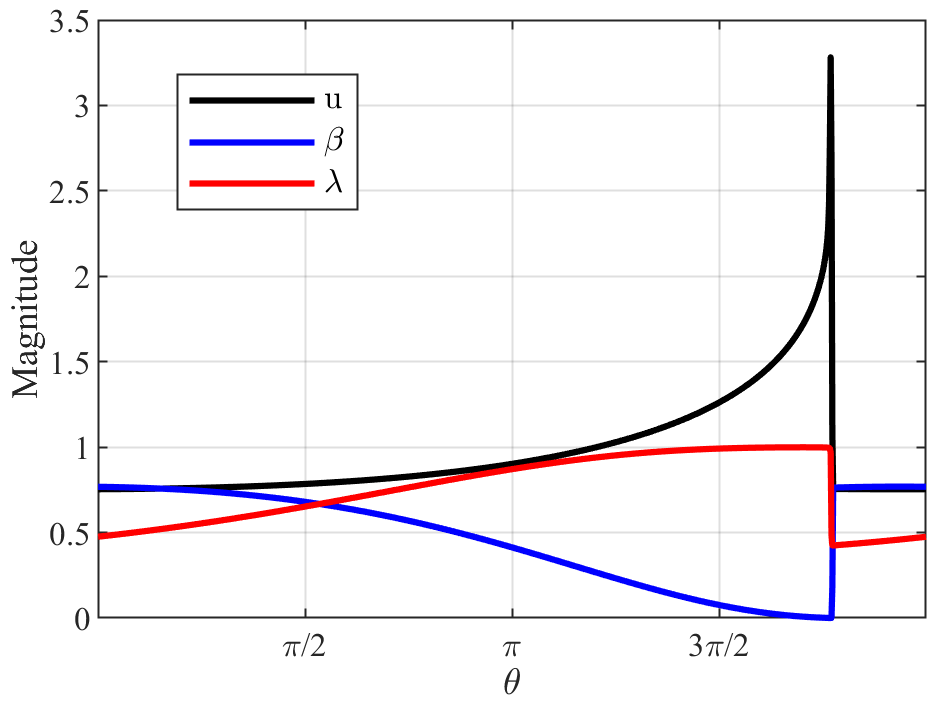}	
		\end{overpic}
		  \caption{Increasing the magnitude of the loss coefficient $\epsilon$ from 0.11 to 0.3 increases the traveling wave speed to 117\% (up from 74\%) of the CJ value referenced to an ambient state of $u_0 = 0$. The simulation is otherwise identical to that of Fig. \ref{fig:pulses}.}
		\label{fig:fast}
\end{figure}

To exemplify this phenomena, Fig. \ref{fig:waveform} includes snapshots in time of the waveforms within the domains of the experiment and a simulation of the model system. Once the para-wave deflagration preceding the detonation can self-steepen, a new wave is nucleated and begins the mode-locking process. With an additional wave, the state of the domain is elevated and para-wave deflagration is exacerbated. Although the wave speeds before and after bifurcations in the model system are comparable (on the order of 10\% jumps in velocity), the developed speeds are the manifestation of both changes in wave amplitudes and combustor states. Therefore, to increase wave speeds and proportion of heating via detonation (compared to deflagration) is analogous to increasing the strength of the restoring force. For example, increasing the restoring force coefficient from $\epsilon = 0.11$ to $0.3$ of the simulation in Fig. \ref{fig:pulses} results in a single wave traveling at 117\% of the CJ speed of the propellant (compared to two waves traveling at 74\% of the CJ speed). The waveform is shown in Fig. \ref{fig:fast}. Note that this is not an over-driven detonation but rather a reference to the CJ wave with a non-elevated ambient state of the domain ($u_0 = 0$).

\section{Conclusion} \label{section:conclusion}

The significance of the proposed model is twofold. First, although we claim no engineering predictive capabilities, our model does relate the dominant physics of gain depletion, gain recovery, and energy dissipation of rotating detonation waves in a simple mathematical framework that recovers, qualitatively, the nonlinear dynamics and bifurcation structure of the waves. This work allows for the immediate analysis of the proposed model to derive stability criteria for RDEs, provide insight into the physical processes behind the rich dynamics of the detonation waves, and aid in the design of future engines. Second, the experimental observations and model extend the well-established physical phenomenon of mode-locking to rotating detonation waves. The energy balance in the RDE combustion chamber is generic, producing mode-locked states that interact through the global gain dynamics.  These dominant balance physics are also observed in well-established laser systems where an analysis of the energy balance produces the global bifurcation structures~\cite{li2010geometrical}.

\begin{acknowledgments}
The authors acknowledge sponsorship under the Air Force Office of Scientific Research (AFOSR) grant FA 9550-18-1-9-007 and Office of Naval Research funding document N0001417MP00398. JNK acknowledges support from AFOSR grant FA9550-17-1-0329.
\end{acknowledgments}

\bibliography{pre_detonation}

\end{document}